\documentclass[aps,pra,twocolumn,superscriptaddress,shownopacs]{revtex4}

\usepackage{graphicx}
\usepackage{hyperref}
\usepackage{amsmath}
\usepackage{epstopdf}
\newcommand{\ket}[1]{\vert #1 \rangle}

\begin{document}

\title{Remote creation of hybrid entanglement \\between particle-like and wave-like optical qubits}

\author{Olivier Morin}
\affiliation{Laboratoire Kastler Brossel, Universit\'{e}
Pierre et Marie Curie, Ecole Normale Sup\'{e}rieure, CNRS, 4 place
Jussieu, 75252 Paris Cedex 05, France}
\author{Kun Huang}
\affiliation{Laboratoire Kastler Brossel, Universit\'{e}
Pierre et Marie Curie, Ecole Normale Sup\'{e}rieure, CNRS, 4 place
Jussieu, 75252 Paris Cedex 05, France}
\affiliation{State Key Laboratory of Precision Spectroscopy, East China Normal University, Shanghai 200062, China}
\author{Jianli Liu}
\affiliation{Laboratoire Kastler Brossel, Universit\'{e}
Pierre et Marie Curie, Ecole Normale Sup\'{e}rieure, CNRS, 4 place
Jussieu, 75252 Paris Cedex 05, France}
\author{Hanna Le Jeannic}
\affiliation{Laboratoire Kastler Brossel, Universit\'{e}
Pierre et Marie Curie, Ecole Normale Sup\'{e}rieure, CNRS, 4 place
Jussieu, 75252 Paris Cedex 05, France}
\author{Claude Fabre}
\affiliation{Laboratoire Kastler Brossel, Universit\'{e}
Pierre et Marie Curie, Ecole Normale Sup\'{e}rieure, CNRS, 4 place
Jussieu, 75252 Paris Cedex 05, France}
\author{Julien Laurat}
\email{julien.laurat@upmc.fr}
\affiliation{Laboratoire Kastler Brossel, Universit\'{e}
Pierre et Marie Curie, Ecole Normale Sup\'{e}rieure, CNRS, 4 place
Jussieu, 75252 Paris Cedex 05, France}

\date{\today} 
\maketitle
 
\textbf{The wave-particle duality of light has led to two different encodings for optical quantum information processing. Several approaches have emerged based either on particle-like discrete-variable states, e.g. finite-dimensional quantum systems, or on wave-like continuous-variable states, e.g. infinite-dimensional systems. Here, we demonstrate the first generation of entanglement between optical qubits of these different types, located at distant places and connected by a lossy channel. Such hybrid entanglement, which is a key resource for a variety of recently proposed schemes, including quantum cryptography and computing, enables to convert information from one Hilbert space to the other via teleportation and therefore connect remote quantum processors based upon different encodings. Beyond its fundamental significance for the exploration of entanglement and its possible instantiations, our optical circuit opens the promises for heterogeneous network implementations, where discrete and continuous-variable operations and techniques can be efficiently combined.}\\

The discrete \cite{Kok07} and the continuous-variable \cite{braunstein03} approaches to linear optical quantum computing and quantum communication \cite{ralphpryde,Obrien} rely on different physical states for their implementation. The first one involves single-photons \cite{KLM}, and the photonic qubits live in a two-dimensional space spanned for example by orthogonal polarizations or the absence or presence of a single-photon, as expressed by $c_0|0\rangle+c_1|1\rangle$. In the continuous alternative, the encoding is realized in the quadrature components of a light field, in an inherently infinite-dimensional space, and the qubits, also sometimes called \textit{qumodes} \cite{vanLoock}, can be implemented for instance as arbitrary superpositions of classical light waves with opposite phases \cite{Jeong02,Ralph03}, $c_0|\alpha\rangle+c_1|-\alpha\rangle$, where $|\alpha\rangle$ is a coherent state with a mean photon number $|\alpha|^2$. 

In parallel to the demonstration of groundbreaking experiments with single-photons, coherent state superpositions also spurred a considerable theoretical and experimental activity over the last years as reminiscent of the Schr\"odinger cat state but also as the main off-line resource for investigating continuous-variable-based protocols. Quantum repeater architectures using this paradigm have been proposed \cite{Sangouard10,Brask10} and there is now a variety of schemes for quantum computing using such a computational basis, including fault tolerant operations with coherent states of moderate amplitude $|\alpha|\sim 1.2$ \cite{Lund08}. 

Both encodings have their own advantages and drawbacks \cite{Park10}. Continuous-variables can benefit from unconditional operations, high detection efficiencies, unambiguous state discrimination and more practical interfacing with conventional information technology. It is however well-known that they suffer from a strong sensitivity to losses and intrinsically limited fidelities. On the other side, discrete-variable approaches can achieve close to unity fidelity but usually at the expense of probabilistic implementations. Combining the two, i.e. achieving hybrid architectures \cite{vanLoock}, may offer serious advantages \cite{Lee13,Morin13}. Some operations might indeed better take advantage of the CV toolbox, while others might be more efficient within the DV framework. In this endeavor, transferring information between the two encodings is a crucial requirement. 

\begin{figure*}[t!]
\centerline{\includegraphics[width=0.96\textwidth]{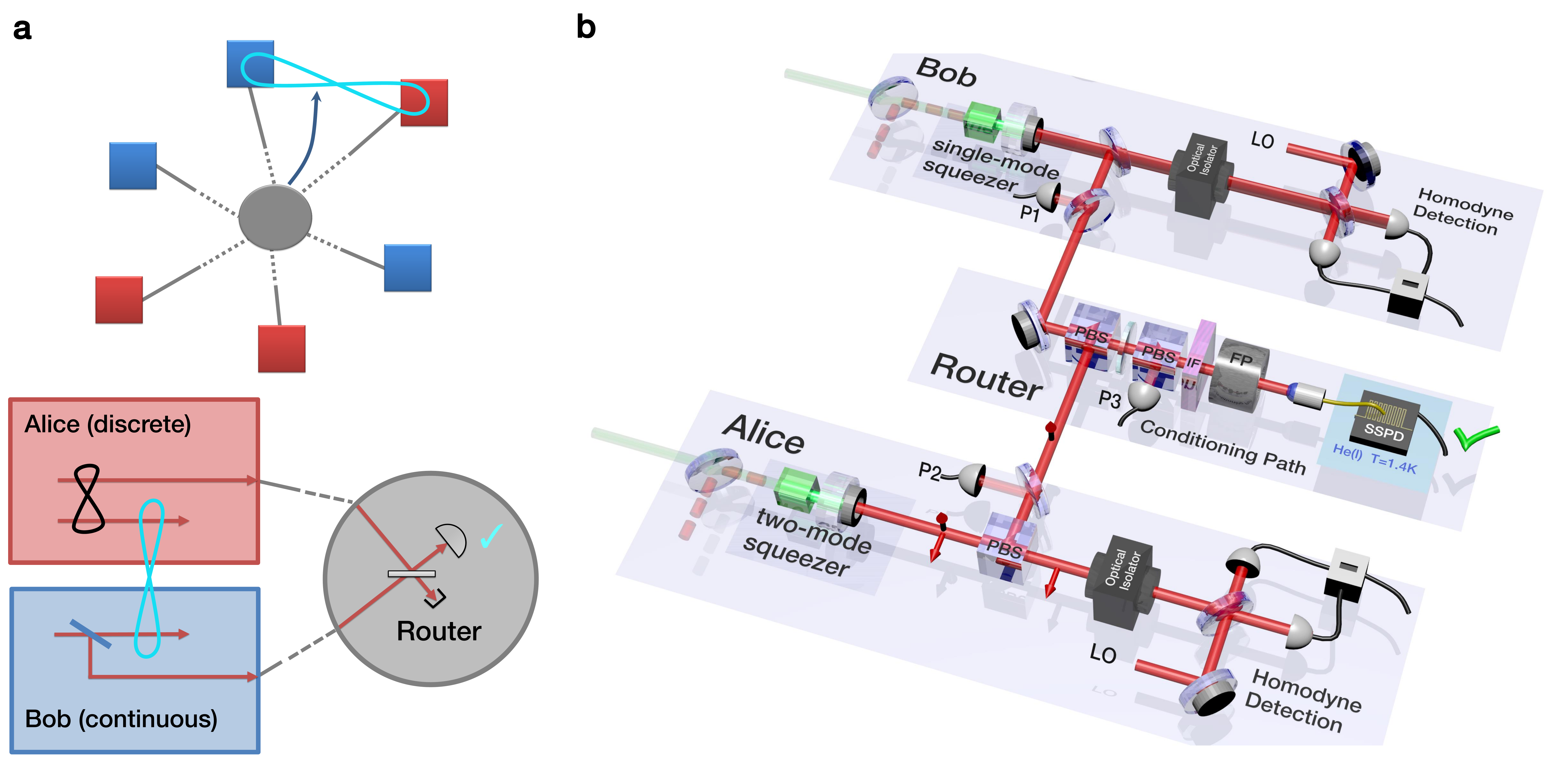}}
\caption{\textbf{Measurement-induced hybrid entanglement.} \textbf{a}, Distant nodes of a quantum network can rely on different information encodings, i.e. continuous (CV) or discrete (DV) variables. A Router enables at a distance to establish hybrid entanglement between the nodes. For instance, Alice sends one mode of a weak two-mode squeezed state $\ket{0}\ket{0}+\lambda\ket{1}\ket{1}$ towards the Router while Bob transmits a small part of a cat state $\ket{\alpha}+\ket{-\alpha}$. The two modes interfere in an indistinguishable fashion on a beamsplitter. Each detection event at the output heralds the generation of hybrid entanglement between Alice and Bob, which can be used for further processing or teleportation. \textbf{b}, Experimental setup. Alice and Bob locally generate the required resources by using continuous-wave optical parametric oscillators operated below threshold. A two-mode squeezer and a single mode-squeezer are used respectively on Alice's and Bob's node. A small fraction of Bob's squeezed vacuum, which is a good approximation of an even cat state for $|\alpha|^2\lesssim1$, is tapped (3\%) and mixed at a central station to the idler beam generated by Alice. The resulting beam is then frequency filtered (conditioning path) and detected by a superconducting single photon detector (SSPD). Given a detection event, which heralds the entanglement generation, the hybrid entangled state is characterized by two high-efficiency homodyne detections. Photodiodes P1, P2 and P3 are used for phase control and stabilization. The beamsplitter ratio in the central station enables to choose the relative weights in the superposition. FP stands for Fabry-Perot cavity, IF for interferential filter, PBS for polarizing beamsplitter and LO for local oscillator.}
\label{fig1}
\end{figure*}

A way to realize a mapping between the two encodings may be provided by teleportation using entanglement between particle-like and wave-like qubits, i.e. hybrid entanglement of the form $|+\rangle|\alpha\rangle+|-\rangle|-\alpha\rangle$, where $|\pm\rangle$ refers to the two-level qubit system \cite{Kreis2012}. Such entanglement can also be useful for quantum key distribution protocols and security analysis \cite{Rigas06,Wittmann10}. Moreover, hybrid entanglement is the key resource for the elegant quantum bus approach where direct qubit-qubit interactions are avoided by being mediated by a common qumode \cite{Spiller06,Loock08}, an example of scheme combining both encodings. It is also the main off-line state required for recently proposed schemes of resource-efficient quantum computation with the promise of near-deterministic universal gate operations \cite{Lee13}. 

 In the present work, in contrast to proposals based on a daunting dispersive light-matter interaction \cite{LoockPRL06} or Kerr non-linearities between single-photon and coherent states (Ref. \cite{Jeong05a} and references therein), we propose and implement a measurement-induced generation of such an entangled state at a distance. Similar to the Duan-Lukin-Cirac-Zoller protocol in the discrete variable regime \cite{DLCZ}, or the remote generation of quasi-Bell states in the continuous variable regime \cite{Ourjoumtsev09}, our scheme relies on a probabilistic preparation heralded by the detection of a single-photon in an indistinguishable fashion. The fragile components remain local and only single-photons propagate between the two distant nodes. A lossy channel affects in this way the count rate but not the fidelity of the resulting state.

\begin{figure*}[t!]
\centerline{\includegraphics[width=0.95\textwidth]{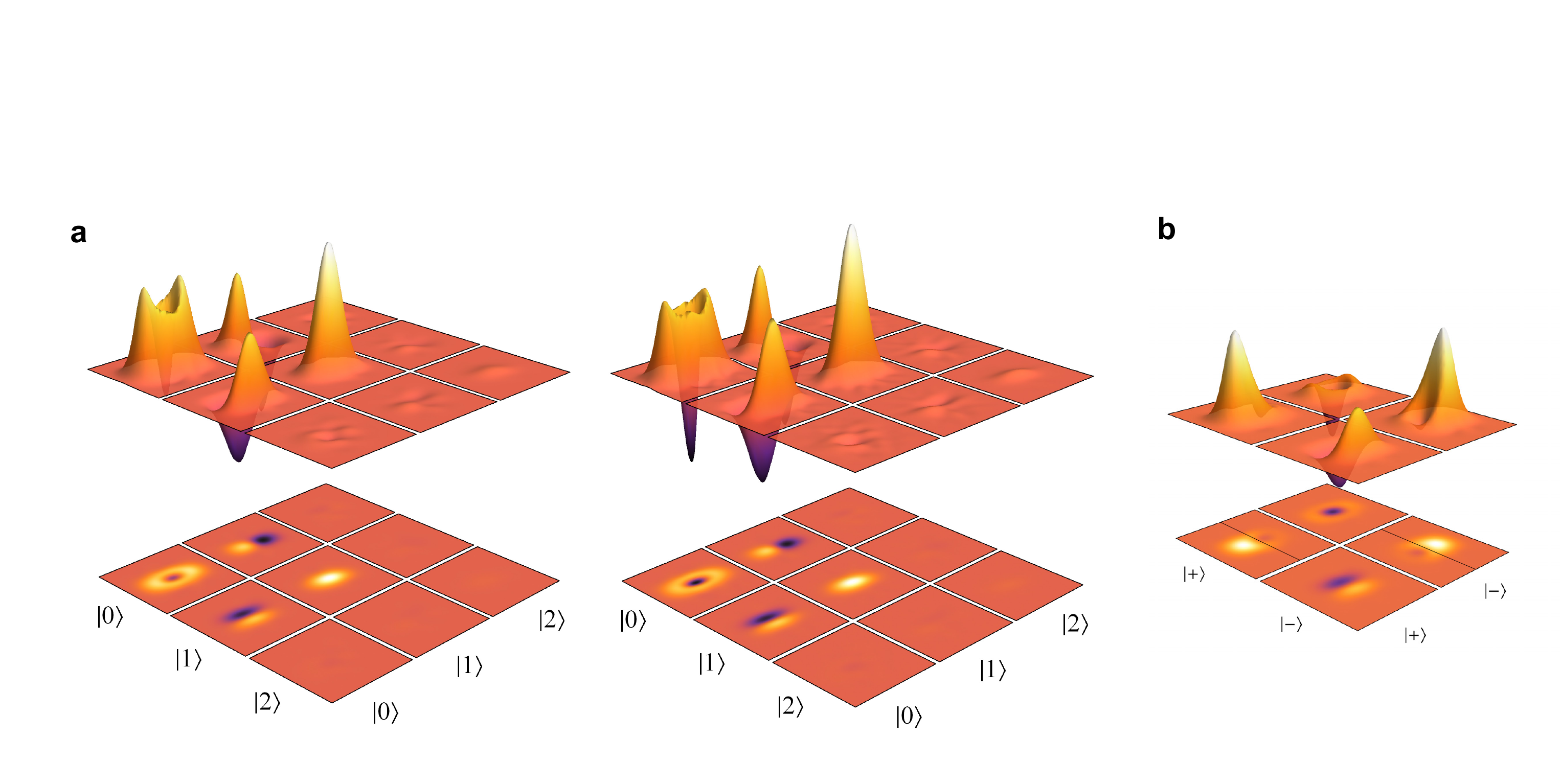}}
\caption{\textbf{Experimental quantum state tomography.} The relative phase is set to $\varphi=\pi$ and the beamsplitter ratio in the central station is adjusted to generate a maximally entangled state, i.e. with equal weights. \textbf{a}, Wigner functions associated with the reduced density matrices $\langle k |\hat{\rho}| l \rangle$ with $k,l$ $\in$ $\{0,1,2\}$, without and with correction for detection losses ($\eta=85\%$). The components with $k\neq l$ being not hermitian, the corresponding Wigner functions are not necessarily real, but conjugate. The plot gives therefore the real part in the back corner ($k<l$) and the imaginary part in the front corner ($k>l$). \textbf{b}, Wigner functions associated with the reduced density matrices $\langle k |\hat{\rho}| l \rangle$ with $k,l$ $\in$ $\{+,-\}$, corrected for detection losses. $|+\rangle$ and $|-\rangle$ stand respectively for the rotated basis $(|0\rangle+|1\rangle)/\sqrt{2}$ and $(|0\rangle-|1\rangle)/\sqrt{2}$. The higher photon number terms are not represented anymore as they are negligible, as shown in \textbf{a}. Corrected for detection losses, the fidelity is 77$\pm$3\% with the targeted state with $\varphi=\pi$ and $|\alpha|=0.9$.}
\label{fig2}
\end{figure*}

The optical circuit is illustrated in Fig. \ref{fig1}a. Alice and Bob (denoted A and B in the following), who are using respectively discrete variable (DV) and continuous-variable (CV) encodings for information processing, are willing to establish hybrid entanglement. In order to establish this internode connection, they prepare locally two non-classical light fields:  a two-mode squeezed state in the very low gain limit on Alice's side, $|\psi\rangle_{si}=|0_s,0_i\rangle+\lambda|1_s,1_i\rangle+O(\lambda^2)$ where $s$ and $i$ stand for the signal and idler modes and a single-mode even cat state on Bob's node, $|\textrm{cat}_+\rangle=|\alpha\rangle+|-\alpha\rangle$. A small fraction of the cat state is then tapped off and transferred to a router station, where it can be superposed on a tunable beamsplitter, in an indistinguishable way, with the idler mode of Alice's state. Conditioned on the detection of a single-photon at the output, and with a beamsplitter ratio adjusted to balance the two contributions, the resulting state is a maximally entangled state for any values of $|\alpha|$, i.e. containing 1 ebit. In the ideal case it can be written as:
\begin{equation}
|\Psi\rangle_{AB}=|0\rangle_A |\textrm{cat}_-\rangle_B +e^{i\varphi}|1\rangle_A |\textrm{cat}_+\rangle_B,
\label{ideal}
\end{equation}
where $\varphi$ is the overall relative phase for the triggering modes, which can be controlled and adjusted, and $|\textrm{cat}_-\rangle=|\alpha\rangle-|-\alpha\rangle$ denotes an odd cat state. 

This  generation procedure can be understood in the following way: a detection event heralds either the subtraction of a single-photon from the even cat state, resulting in a parity change and leaving Alice's signal mode in a thermal state very close to the vacuum state, or the detection of a single-photon in the idler mode, resulting in projecting the signal into a single-photon state and leaving unchanged the initial cat state.

The resulting entangled state can also be written using for Alice the rotated qubit basis $\left\{|+\rangle=(|0\rangle+|1\rangle)/\small{\sqrt{2}},|-\rangle=(|0\rangle-|1\rangle)/\sqrt{2}\right\}$ as: 
\begin{equation}
|\Psi\rangle_{AB}=|+\rangle_A |\alpha\rangle_B +e^{i\varphi'}|-\rangle_A |-\alpha\rangle_B.
\label{idealrotated}
\end{equation} 
The normalizations are omitted here and this rewriting from eq. (\ref{ideal}) is valid when the two coherent states are approximately orthogonal. Let us note however that this approximation is already good for moderate values of $|\alpha|$ \cite{Ralph03,Lund08}. Indeed, an amplitude $|\alpha|= 1$ gives an overlap $|\langle\alpha|-\alpha\rangle|^2=e^{-4\alpha^2}\sim0.02$.

This state directly enables the teleportation of a qubit encoded in the $\left\{|+\rangle,|-\rangle\right\}$ basis to the coherent state computational basis $\left\{|\alpha\rangle,|-\alpha\rangle\right\}$. It also refers to the spirit of the Schr\"odinger Gedankenexperiment where the two classical states are entangled with a microscopic degree of freedom. Let us note also that a Hadamard gate, which can be performed with non-gaussian ancilla and projective measurements \cite{Marek2010}, would enable to convert this state into the state $|0\rangle_A |\alpha\rangle_B +|1\rangle_A |-\alpha\rangle_B$.

In the present work, the non-classical fields are generated at 1064 nm with continuous-wave optical parametric oscillators (OPO) operated below threshold (see appendix), as illustrated in Fig. \ref{fig1}b. Importantly, these sources enable the generation of quantum states in a very well-defined spatio-temporal mode due to the cavity \cite{Morin2013}. On Bob's site, a type-I OPO is used to generate a single-mode squeezed vacuum with 3 dB noise reduction below shot noise level. For $|\alpha|^2\lesssim1$, this state has a close to unity fidelity with an even cat state $|\textrm{cat}_+\rangle$. A small fraction, $R=3\%$, of the light is tapped off via a beamsplitter. Subtracting a single-photon from this state results in the generation of a odd cat state \cite{Ourjoumtsev06,Polzik,Wakui,Noriuki}. On Alice's side, the required two-mode squeezed vacuum is generated by a type-II frequency-degenerate OPO. At the output, the orthogonally polarized signal and idler modes are spatially separated via a polarizing beam splitter. The device is operated very far below threshold (around 100 times below) in order to limit the multi-photon component to a few percents \cite{Morin2012}. The tapped mode and the idler mode are then brought to interfere. Before detection, frequency filtering elements are needed to remove the non-degenerate modes emitted by the OPOs. Finally, the filtered mode is detected by a superconducting single-photon detector (SSPD, Scontel) working at cryogenic temperature. The very low dark noise (below 1Hz) avoids false detection events, a crucial feature to achieve high-fidelity in the state generation \cite{Dauria,Dauria2}.

To achieve entanglement, various parameters have to be strictly controlled. First, the superposed beams must be indistinguishable. This stringent condition requires in particular to match the bandwidth of the two OPOs. Starting from similarly-built OPOs, a fine tuning of the cavity lengths is performed by adjusting the temporal modes in which the conditional states are emitted when operated separately. Second, the relative phase $\varphi$ has to be kept constant \cite{Kimble}. The different phases in the experiment are therefore controlled and actively stabilized by using auxiliary weak beams injected into both OPOs (see appendix). For this purpose, the experiment is conducted in a cyclic fashion: 50 ms are used for phase locking and the data acquisition then starts for the next 50 ms with seed beams off.

The heralded state $\hat{\rho}$ is characterized by a two-mode quantum tomography performed with two high-efficiency homodyne detections ($\eta=85\%$), one on each node. 200,000 data points are recorded with equally distributed choice of quadratures. The two-mode density matrix of the state is then reconstructed via a maximum likelihood algorithm \cite{Lvovsky}. This new kind of hybrid state, which is composed of a discrete mode and a continuous one, raises the question of how to represent it in a visual and illustrative manner. We have chosen as a convenient representation to display in a matrix form the Wigner functions (well-adapted to continuous-variable states) associated with the reduced density matrices $\langle k |\hat{\rho}| l \rangle$, where $|k\rangle$,$|l\rangle$ stand for the discrete qubit states. 

The experimental results are given in Fig.\ref{fig2}a without and with correction for detection losses, for a phase set to $\varphi=\pi$ and a beamsplitter ratio tuned to balance the detection probability from each nodes. This figure first confirms that the discrete mode is contained into the qubit subspace spanned by $\{|0\rangle$,$|1\rangle\}$. Higher photon number components are indeed limited to 2\%. In this subspace, the two first diagonal elements, namely the projection $\langle 0 |\hat{\rho}| 0 \rangle$ and $\langle 1 |\hat{\rho}| 1 \rangle$, correspond respectively to a photon-subtracted squeezed state and to a squeezed state. The non-zero off-diagonal terms witness the coherence of the superposition. The generated state can also be represented using as another projection basis the rotated one $\left\{|+\rangle=(|0\rangle+|1\rangle)/\sqrt{2},|-\rangle=(|0\rangle-|1\rangle)/\sqrt{2}\right\}$ (Fig. \ref{fig2}b). As it can be clearly seen from the contour plots, the two projections $\langle + |\hat{\rho}| + \rangle$ and $\langle - |\hat{\rho}| - \rangle$ exhibit an opposite displacement in phase space, corresponding with large fidelity to the two states $|\alpha\rangle$ and $|-\alpha\rangle$. Corrected for detection losses, we obtain a fidelity 77$\pm$3\% with the targeted state with $\varphi=\pi$ and $|\alpha|=0.9$. The demonstrated size is already compatible with the value $|\alpha|\sim1$ shown as the optimal value in recent proposals of ressource-efficient operations with hybrid qubits \cite{Lee13}.

To quantitatively assess the generated entanglement, we also compute the negativity \cite{Vidal02} given by $\mathcal{N}=\left(||\rho^{T_A}||_1-1\right)/2$, where $\rho^{T_A}$ stands for the partial transpose of the two-mode density matrix $\rho$ with respect to mode $A$. This quantity reaches 0.5 for the ideal maximally-entangled state. Experimentally, $\mathcal{N}=0.26\pm0.01$ is obtained without correction for detection losses and $\mathcal{N}=0.37\pm0.01$ when corrected, demonstrating the hybrid entanglement remotely prepared and its suitability for hybrid teleportation \cite{Park12}. The heralding rate is equal to 30 kHz, limited by an overall loss in the conditioning path equal to 97\% \cite{Morin2012}. This lossy channel would be equivalent to 75 km of fibre at telecom wavelength. This value confirms the reliability of our method to establish entanglement connection on long distances. 

Experimental imperfections can be summarized on both nodes by an effective local efficiency, $\eta_A$ and $\eta_B$, which mostly arises here from transmission losses, finite detection efficiency and escape efficiency of the OPO given by $T/(T+L)$ where $T$ is the transmission of the output coupler and $L$ the intracavity losses. Dark counts are negligible in our experiment. By using the value of the Wigner function at the origin for the states generated independently, these efficiencies are estimated to be $\eta_A=76\pm2\%$ and $\eta_B=71\pm2\%$. These values are in agreement with the observed negative value at the origin of the Wigner function for the odd cat state $\langle 0 |\hat{\rho}| 0 \rangle$ given in Fig. \ref{fig2}a, $W_0=-0.14\pm0.01$ (not corrected, ideally $-1$). Let us note that both local efficiencies contribute in this resulting value. Indeed, projecting on Alice's side translates into adding an extra vacuum contribution on Bob's side resulting from her non-unity local efficiency. Achieving negativity without correction is thus a difficult task here and constitutes a notable feature of our work.  

\begin{figure*}[htpb!]
\centerline{\includegraphics[width=0.98\textwidth]{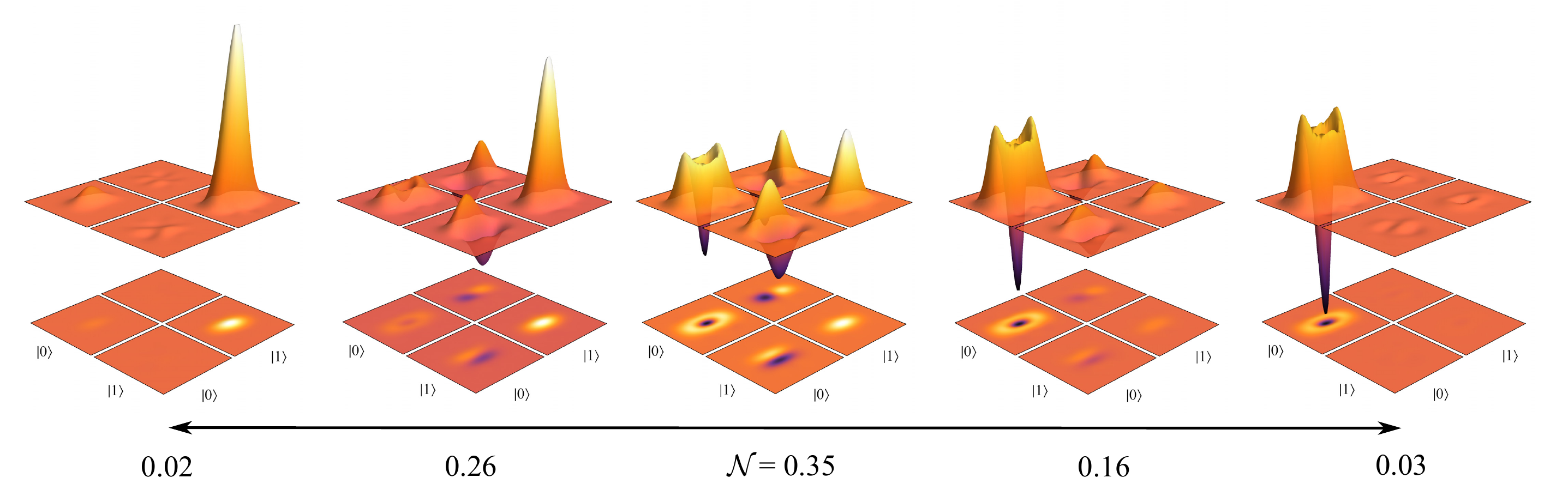}}
\caption{\textbf{Experimental quantum states from separable to maximally entangled.} The relative phase is set to $\varphi=0$ and the beamsplitter ratio at the central station is tuned. The blocks provide the Wigner functions associated with the reduced density matrices $\langle k |\hat{\rho}| l \rangle$ with $k,l$ $\in$ $\{0,1\}$. The higher photon number terms are not represented as they are negligible. For each generated state, the negativity $\mathcal{N}$ (equal to 0.5 in the ideal case) is computed from the full two-mode density matrix, showing the transition from separable to maximally entangled state and back to separable.}
\label{fig3}
\end{figure*}

The accurate control on the experimental parameters achieved in our implementation also enables the complete engineering of the hybrid state by choosing the relative phase $\varphi$ and the superposition weights. Figure \ref{fig3} provides the example of states with $\varphi=0$, i.e. opposite to the one used in Fig.\ref{fig2}, and for different ratio of the beamsplitter used for the mixing. The two extreme states indeed result from heralding events coming only from Alice's node or Bob's node. The figure in the middle provides the balanced case, which is very similar to the one provided in Fig.\ref{fig2}a but with an opposite phase, as can be seen in the off-diagonal terms. The two other blocks give examples of intermediate ratios, showing the building up of the coherences. The negativity is provided in each case, showing the transition from separability to entanglement and back.

As can be seen in Fig.\ref{fig2}b in the rotated basis, the projected states are not completely round as expected for coherent states. This feature comes from the initial approximation consisting in starting with a squeezed vacuum in the experimental protocol. The maximum achievable fidelity with the targeted hybrid state is therefore 94\% in this case. To go beyond this demonstrated result in future extensions, Bob can perform a local single-photon subtraction to initially prepare an odd cat state. This preparation will directly enable to increase $|\alpha|$ above 1 and the achievable fidelity close to unity. Higher amplitudes can be obtained by additional photon subtractions \cite{Molmer}. Our optical circuit is well-suited for these operations and provides a platform for subsequent experiments. Higher values of squeezing with high purity will be necessary, and are readily available given our OPO escape efficiency \cite{jove}. Efficient photon detectors are also required for these cascaded detections and can be provided by the new generation of superconducting devices working in the near infrared \cite{Marsili}. In a different framework, these subsequent works will also enable the study of squeezing-induced micro-macro states as recently proposed in Ref. \cite{Andersen2013} where the phase-space distance can be varied, in contrast to recently reported results based on displaced single-photon entanglement \cite{Bruno2013,Lvovsky2013}.

In summary, we have achieved entanglement between two remote nodes that are using different information encodings. Living in Hilbert space of different dimensionality, the two parties establish heralded hybrid entanglement, which enables for instance to map discrete qubits onto coherent state ones. The work presented here constitutes the first demonstration of such hybrid entanglement enabling to link computational basis of different nature. This possibility, in combination with further works on high-fidelity quantum state engineering, provides a new resource for optical hybrid architecture and quantum network operation based on heterogeneous systems. \\

\acknowledgements 
We thank N. Sangouard for interesting discussions and V. D'Auria and F.A.S. Barbosa for their valuable contributions in the early stage of the experiment. This work is supported by the ERA-Net CHIST-ERA (QScale) and by the European Research Council (ERC) Starting grant HybridNet. K.H. acknowledges the support from the Foundation for the Author of National Excellent Doctoral Dissertation of China (PY2012004) and the China Scholarship Council. C.F. and J.L. are members of the Institut Universitaire de France.

\appendix
\section{Broadband optical parametric oscillators} The two triply-resonant optical parametric oscillators are pumped by a continuous-wave frequency-doubled Nd:YAG laser (Innolight GmbH) and operated below threshold. The first one is a two-mode squeezer based on a type-II phase-matched KTP crystal while the second one is a single-mode squeezer using a type-I PPKTP crystal. Both crystals (Raicol Crystals) are 10 mm long and the OPOs are made of a semimonolithic linear cavity: the input mirror is coated on one face of the crystal (high reflection at 1064 nm and a transmission equal to 5\% at 532 nm) and the output mirror, with a 38 mm radius of curvature, is highly reflective for the pump and with a reflection for the infrared equal to 90\%. The cavities are locked on resonance by the Pound-Drever-Hall technique with a 12 MHz phase modulation on the pump. In order to achieve a high indistinguishability in the heralding event, the bandwidth of the OPOs are precisely matched by adjusting their lengths. The type-II OPO is pumped very far below threshold in order to limit the multi-photon components when a photon is detected from the idler mode and, in the same way, limit the contamination of the vacuum state by higher terms when the detected photon originated from the other OPO. In both cases, the ratio is the same giving by the parametric gain. \\

\section{Frequency filtering in the conditioning path} In order to remove the non-degenerate modes emitted by the optical parametric oscillators, and consequently only detect heralding single-photons at the carrier frequency, a two-stage filtering is applied. An interferential filter (Barr Associates) with a bandwidth equal to 0.5 nm is therefore associated with a home-made linear Fabry-Perot cavity (0.45 mm long and with a finesse equal to 1000), which has a free spectral range of 330 GHz and a bandwidth of 320 MHz (6 times larger than the one of the optical parametric oscillators). The unwanted modes are rejected below 0.3 \%. \\

\section{Control of the phases} The relative phase $\varphi$ in the generated state, which critically has to be kept constant to achieve entanglement, is determined by the optical paths after the OPOs but also by the phases of each OPO pump field. In order to control these phases, a very weak seed beam is injected into each OPO. These beams are amplified or de-amplified depending on the relative phase between the seed and the pump. To set this phase, a small part of each beam is measured (photodiodes P1 and P2 in Fig.1) and the intensity is locked at a constant level. The second step consists in locking the relative phase where the tapped modes are combined (Photodiode P3). Additionally, the interference between the seed and the local oscillator on each site gives access to the phase of the measured quadrature for the subsequent quantum state tomography. \\

\end{document}